\documentclass[fleqn,usenatbib,useAMS]{mnras}
\usepackage{newtxtext,newtxmath}
\usepackage{graphicx}
\usepackage[normalem]{ulem}

\title[Dwarf Galaxies without Dark Matter]{Dwarf Galaxies without Dark Matter: constraints on  Modified Gravity}

\author[Khalifeh \& Jimenez]{Ali Rida Khalifeh$^{1,2}$\thanks{ark93@icc.ub.edu}, Raul Jimenez$^{2,3}$\thanks{raul.jimenez@icc.ub.edu}\\
$^1$Instituto de Ciencias del Cosmos, University of Barcelona, Marti  i Franques, 1, E-08028 Barcelona, Spain.\\
$^2$Dept. de  Fisica Cuantica y Astrofisica, University of Barcelona, Marti  i Franques 1, E-08028 Barcelona, Spain.\\
$^3$Instituci\`o Catalana de Recerca i Estudis Avan\c{c}ats, Pg. Lluis Companys 23, Barcelona, E-08010, Spain.\\
}

\begin{document}

\voffset=-0.25in 

\maketitle

\begin{abstract}
The discovery of $19$ dwarf galaxies without dark matter provides, counter-intuitively, strong support for the $\Lambda$CDM standard model of cosmology. Their presence is well accommodated in a scenario where the dark matter is in the form of cold dark particles.  However, it is interesting to explore quantitatively what is needed from modified gravity models to accommodate the presence of these galaxies and what extra degree of freedom is needed in these models.
To this end, we derive the dynamics at galaxy scales (Virial theorem) for a general class of modified gravity models. We distinguish between theories that satisfy the Jebsen-Birkhoff theorem, and those that don't. Our aim is to develop tests that can distinguish whether dark matter is  part of the theory of gravity or a particle. The 19 dwarf galaxies discovered provide us with a stringent test for models of modified gravity. Our main finding is that there will always be an extra contribution to the Virial theorem coming from the modification of gravity, even if a certain galaxy shows very small, if not negligible, trace of dark matter, as has been reported recently.  Thus, if these and more galaxies are confirmed as devoid (or negligible)  of dark matter, while other similar galaxies have abundant dark matter, it seems interesting to find modifications of gravity to describe dark matter. Our result can be used by future astronomical surveys to put constraints on the parameters of modified gravity models at astrophysical scales where dark matter is described as such.
\end{abstract}


\section{Introduction}\label{sec:intro}
The existence of Dark Matter(DM) has been demonstrated observationally in many occasions. Initially, at astrophysical scales, the Virial mass of the Coma galaxy cluster was found by Zwicky~\citep{Zwicky:1933gu,Salucci:2007tm} to be 500 times larger than the observed one. Later, the flat behavior of stars' velocity curves in the outskirts of spiral galaxies was also another proof of the existence of additional unobserved matter~\citep{Sofue:1999jy,Sofue:2000jx,Persic:1995ru}. Moreover, from the Cosmic Microwave Background structure at cosmological scales, there is a clear evidence for DM~\citep{Bennett:2003bz,Komatsu:2008hk}; see~\cite{Bertone:2016nfn,Freese:2008cz} and references within for a detailed overview of DM. However, the nature of this unobserved entity is still an open question and an active field of research. In the context of the General Theory of Relativity(GR), this phenomenon is described by adding cold particles, that is pressurless non-relativistic ones, to the energy content of the universe. To describe the theory, given a metric of spacetime $g_{\mu\nu}$, one would write an action of the form:
\begin{equation}
S=\int d^4x \sqrt{-g}\bigg[\frac{M_p^2}{2}R(g_{\mu\nu})+\mathcal{L}_{m,r}(g_{\mu\nu},\psi_m,\psi_r)+\mathcal{L}_{DM}\bigg]
\label{eq:FirstAction}
\end{equation}
where $g$ is the determinant of $g_{\mu\nu}$, $R$ is the Ricci scalar, the trace of the Ricci Tensor $R_{\mu\nu}$, $M_p^2=(8\pi G)^{-1}$ is the reduced Planck mass (in units for which the reduced Planck constant $\hbar$ and the speed of light $c$ are 1), $\mathcal{L}_{m,r}$ is the Lagrangian density of matter and radiation, given as a function of the metric and the corresponding fields, $\psi_m$, $\psi_r$ and finally $\mathcal{L}_{DM}$ is the Lagrangian density of DM particles. By setting the variation of~\eqref{eq:FirstAction} with respect to the metric to 0, we get the Einstein equations of motion:
\begin{equation}
\frac{1}{\sqrt{-g}}\frac{\delta S}{\delta g^{\mu\nu}}=0\Rightarrow R_{\mu\nu}-\frac{1}{2}g_{\mu\nu}R=\frac{1}{M_p^2}(T_{\mu\nu}+T_{\mu\nu}^{DM})
\label{eq:Einstein}
\end{equation}
where $T_{\mu\nu}$ is the stress-energy tensor of the baryonic and leptonic matter, as well as radiation, whereas $T_{\mu\nu}^{DM}$ is that of the DM particles. From here one can see how the gravitational phenomena observed (LHS of the above equation) is affected by the presence of DM particles(RHS). The particle nature proposal for DM has presented many candidates beyond the Standard Model of Particle Physics. These include sterile neutrinos~\citep{Dodelson:1993je}, axions~\citep{Visinelli:2009zm,Duffy:2009ig}, and WIMPs(Weakly Interacting Massive Particles), which include the lightest supersymmetric stable particle, the neutralino. For a detailed review on the different particle candidates for DM, see~\cite{Bertone:2010zza} and~\cite{Profumo:2019ujg}.

Another explanation for these phenomena is to consider a theory of gravity other than GR, which is known as Modified Gravity Theory(MGT). In this context, the gravitational laws of nature have specific geometrical properties that could result in the observed phenomena caused by DM, without the need for adding extra species to the particle content of the universe. For instance, one of the proposed MGTs is called $f(R)$ gravity, where $f(R)$ stands for an arbitrary (in the appropriate units) function of the Ricci scalar $R$. In this MGT, one generalizes the Einstein-Hilbert action to:
\begin{equation}
S_{f(R)}=\int d^4x\sqrt{-g}\bigg[\frac{M_p^2}{2}f(R)+\mathcal{L}_{m,r}\bigg].
\label{eq:Actionf(R)}
\end{equation}
The resulting Einstein equations would take the form:
\begin{equation}
R_{\mu\nu}-\frac{1}{2}g_{\mu\nu}R=\frac{1}{M_{p_{eff}}^2}\bigg(T_{\mu\nu}+\tilde{T}_{\mu\nu}\bigg)
\label{eq:Einsteinf(R)}
\end{equation}
where the effective Planck mass is
\begin{equation}
M_{p_{eff}}^2=M_p^2f'(R)
\label{eq:PlanckMassEffective}
\end{equation}
and $'$ denotes the derivative of a function with respect to its argument. The additional term on the RHS,
\begin{equation}
\tilde{T}_{\mu\nu}=M_P^2\bigg[\frac{f(R)-Rf'(R)}{2}g_{\mu\nu}+\nabla_{\mu}\nabla_{\nu}f'(R)-g_{\mu\nu}\Box f'(R)\bigg]
\label{eq:Ttild}
\end{equation}
can now generate the gravitational phenomena observed associated to DM, but this term is not related to some type of particle, rather to gravity itself. In this way, one can provide an alternative explanation to the existence of DM. Another MGT that has been recently proposed is called Mimetic Dark Matter(MDM)~\citep{chamseddine:mimeticdm,Golovnev:lagrangeMultiplier,Barvinsky:ghost}. The original proposal of this work was to rewrite the physical metric in terms of an auxiliary one and the derivative of a scalar field. The resulting equation of motion will resemble~\eqref{eq:Einsteinf(R)} with a different $\tilde{T}_{\mu\nu}$ and $M_{p_{eff}}^2$, but can describe the gravitational effects of DM. The model was further developed to incorporate other cosmological phenomena~\citep{chamseddine:mimeticcosmo}, as well as to avoid problems related to defining quantum fluctuations, adiabatic initial conditions and cosmological singularities~\citep{Chamseddine:2016uef,Mirzagholi:2014ifa,Ramazanov:2015pha}. For further analysis and study of the model, see~\cite{ganz:mimeticgw,Ganz:2018mqi,arroja:mimeticcosmoperturbations,arroja:mimeticdisformallagrange,arroja:mimeticlss,Khalifeh:2015tla,Khalifeh:2019zfi}. More recently, the model has been developed to avoid the original singularity of the universe by having a running gravitational constant~\citep{Chamseddine:2019bcn}. For more reviews on MGTs, see~\cite{Clifton:2011jh,Nojiri:2017ncd}.

The main purpose of this letter is to study the DM phenomena at astrophysical scales using the MGT approach. More specifically, we derive the Virial theorem for a general class of MGTs, including the Horndeski model~\citep{Horndeski:1974wa,Kobayashi:2019hrl}, and see where the observed additional Virial mass comes from. We distinguish, however, our derivation for theories that satisfy the Jebsen-Birkhoff Theorem(JBT)~\citep{1923rmp..book.....B,Jebsen} and those that don't, for reasons that will be explained below. We notice here that the additional Virial mass term will exist irrespective of the system under consideration. Therefore, if one wants to associate $\tilde{T}_{\mu\nu}$ to DM, one would be claiming that their effects exist everywhere, by the universality of gravitational interactions. However, one might wonder what if there is a system in which there is no traceable amount of DM, as has been recently observed~\citep{danieli2019tip,van_Dokkum_2019,Danieli_2019,Guo2019wgb,Pina2019rer}. Even though these results are still being further analyzed, we use the possibility of having systems with no traceable amount of DM to put constraints on the parameters of MGTs in general.
\begin{figure}
\centerline{
\includegraphics[width=.5\columnwidth]{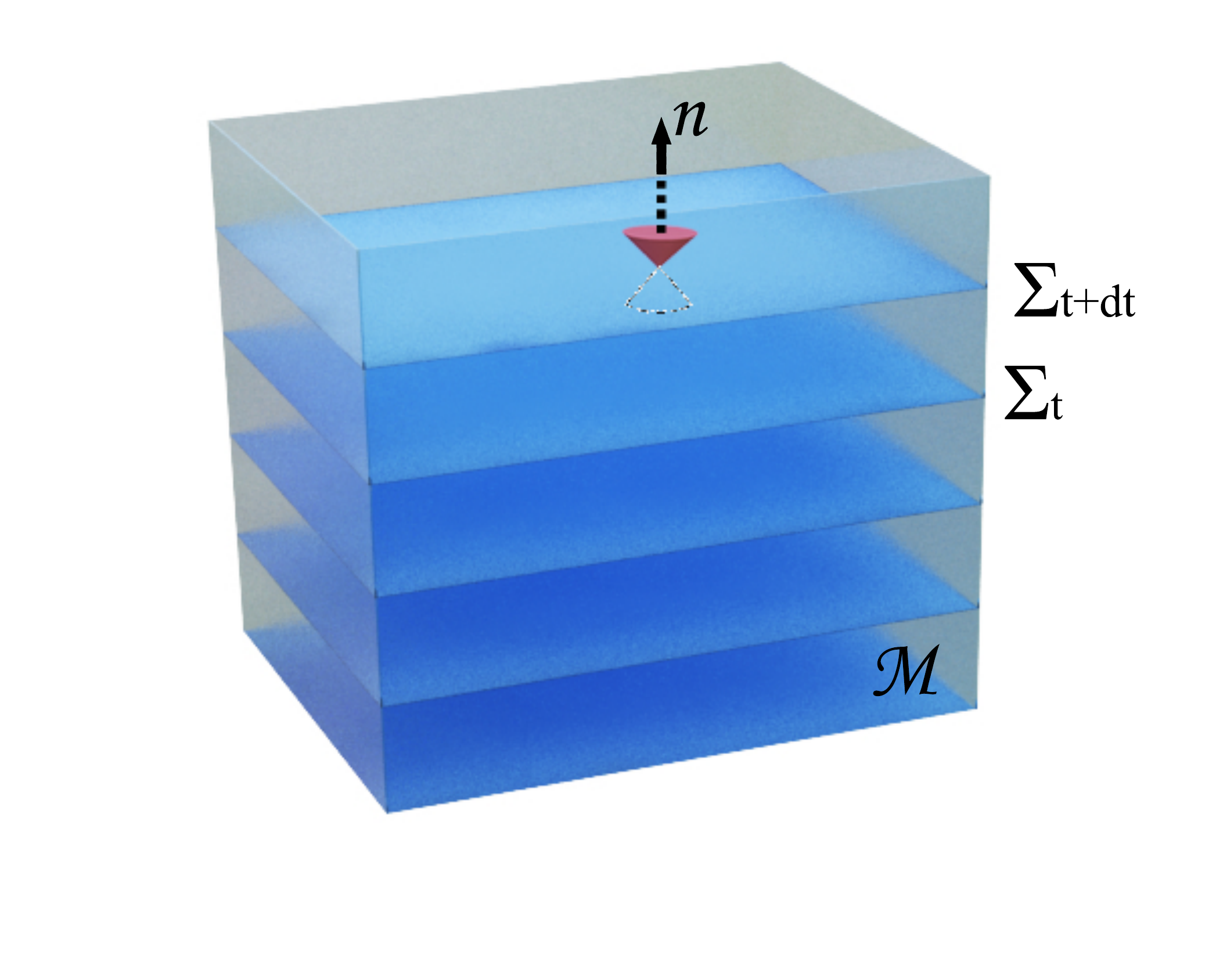}}
\caption{Illustration of foliating a manifold $\mathcal{M}$ with a set of hypersurfaces $\Sigma_t$ in the 3+1 formalism of GR}
\label{fig:ManifoldFoliation}
\end{figure}

\section{Virial Theorem in MGT}\label{sec:VirialTheorem} 
In this section, we derive the Virial Theorem for a class of MGTs that generate equations of motion with the form given in~\eqref{eq:Einsteinf(R)} within the 3+1 formalism of GR~\citep{Arnowitt2008}. A vital element in this derivation is the assumption of asymptotic flatness and stationarity (explained below in more detail), which is valid only if the MGT satisfies the JBT, such as in Brane cosmology or Palatini f(R) gravity (see~\cite{Clifton:2011jh,Faraoni:2010rt,Sotiriou:2008rp,PhysRevD.77.064016} and references therein for more details on theories that do and do not satisfy the JBT). Therefore, the treatment in section~\ref{FirstSection} will be applicable mainly to the former case, while a slight deviation from that will be presented in section~\ref{Violation} for theories that violate the JBT. An alternative derivation of the Virial theorem, using the Lagrangian formalism, will be briefly present in appendix~\ref{Appendix}. This method is applicable to both types of theories described here, for the Virial theorem relies on the collisionless Boltzmann equation, and therefore it's a consequence of stress-energy conservation. This means it should be applicable to any metric theory of gravity~\citep{Schmidt:2010jr}.

\subsection{Theories Satisfying the JBT}\label{FirstSection}

\subsubsection{Formalism}
Consider a stationary and asymptotically flat spacetime\footnote{\label{footnote}
	Mathematically, stationarity means that there exists a time-like Killing vector, at least at spatial infinity, that can be normalized to -1. Asymptotically flat, on the other hand, means two things: First, the $\Sigma_t$s contain a compact region $P$, such that $\Sigma_t-P$ is diffeomorphic to $\Re^3-\{0\}$, where $\Re^3$ is the 3-dimensional real space. Second, one can establish on each $\Sigma_t$ a coordinate system in a way that the components of the metric differ from those of the Minkowski one by $\mathcal{O}(1/r)$ as $r\rightarrow\infty$, where $r$ is the radial distance.}
$\mathcal{M}$ with a metric $g$, and consider foliating $\mathcal{M}$ with a set of space-like hypersurfaces $\Sigma_t$, as illustrated in figure~\ref{fig:ManifoldFoliation}. Moreover, let $n$ be a time-like 4-vector field, orthonormal to the $\Sigma_t$s and directed along increasing time $t$:
\begin{equation}
n_{\alpha}=-Nt_{,\alpha}\Rightarrow n_{\alpha}n^{\alpha}=-1
\label{eq:LapseVector}
\end{equation}
where $_{,\alpha}$ means $\partial/\partial x^{\alpha}$, and $N$ is the strictly positive lapse function. The latter measures the rate of flow of proper time $\tau$ with respect to coordinate time as one moves normally from one $\Sigma_t$ to the next along $n$. Let
\begin{equation}
h_{\alpha\beta}=g_{\alpha\beta}+n_{\alpha}n_{\beta}
\label{eq:ProjectionMetric}
\end{equation}
be the projection tensor orthogonally onto $\Sigma_t$ and, when restricted to $\Sigma_t$, defines the positive definite induced 3-metric by $g$ on $\Sigma_t$. Furthermore, define the shift vector $N^{\alpha}$ as the measure of how much the spatial coordinates shift as they move from one $\Sigma_t$ to the next along $n$:
\begin{equation}
N^{\alpha}=-h^{\alpha}_{\ \beta}\xi^{\beta}
\label{eq:ShiftVector}
\end{equation}
where $\xi^{\alpha}=(\partial/\partial t)^{\alpha}$ is the Killing vector associated with the stationarity of $\mathcal{M}$(see footnote~\ref{footnote}). From these definitions, one can write the explicit components of $n$ and $N$ as:
\begin{eqnarray}
n_{\alpha} & = & (-N,0,0,0); \\ \nonumber
n^{\alpha} & =  & (1/N,N^1/N,N^2/N,N^3/N); \\ \nonumber 
N^{\alpha} & = & (0,N^1,N^2,N^3) 
\label{eq:LapseVectors}
\end{eqnarray}
 and the metric components would be:
\begin{equation}
g_{\mu\nu}dx^{\mu}dx^{\nu}=-(N^2-N_iN^i)dt^2-2N_idtdx^i+h_{ij}dx^idx^j.
\label{eq:metricADMform}
\end{equation}
\\The starting point in deriving the Virial theorem is to contract the Einstein equation~\eqref{eq:Einsteinf(R)} with $h^{\mu\nu}$:
\begin{equation}
R_{\mu\nu}n^{\mu}n^{\nu}-\frac{1}{2}R=\frac{1}{M_{p_{eff}}^2}\big[S_{\mu}^{\ \mu}-\tilde{S}_{\mu}^{\ \mu}\big]
\label{eq:EinsteinEquationContracted}
\end{equation}
where
\begin{equation}
S_{\mu}^{\ \mu}\big(\tilde{S}_{\mu}^{\ \mu}\big)=h^{\mu\nu}T_{\mu\nu}\big(\tilde{T}_{\mu\nu}\big).
\label{eq:TraceEMT}
\end{equation} 
We can now use the Gauss-Codazzi-Mainardi equations, which relate the Ricci tensor of the 4-metric to that of the 3-metric $h_{\mu\nu}$, $^{3}R_{\mu\nu}$, the lapse function $N$ and the extrinsic curvature of $\Sigma_t$, $K_{\mu\nu}$(see~\cite{PhysRevD.48.2635} for more details). The final result would be:
\begin{equation}
\nu^{|i}_{\ \ |i}-\frac{1}{4}{}^{3}R+\nu^{|i}\nu_{|i}-\frac{3}{4}\big(K_{ij}K^{ij}-K^2\big)+(Kn^{\alpha})_{;\alpha}=\frac{1}{2M_{p_{eff}}^2}\big[S_{i}^{\ i}-\tilde{S}_{i}^{\ i}\big]
\label{eq:EndStep1}
\end{equation}
where "$_{|i}$" denotes the covariant derivative with respect to $x^i$ associated with the 3-metric $h$, "$_{;\alpha}$" is the covariant derivative with respect to $x^{\alpha}$ associated with the 4-metric $g$, $\nu=\ln N$ and $K=-n^{\alpha}_{\ \ ;\alpha}$ is the trace of $K_{\mu\nu}$. Now that we have done the first step, we can proceed to the second one, which is to integrate this result over space.

\subsubsection{Step 2: Integration Over Space}\label{sec:Step2}
Integrating~\eqref{eq:EndStep1} over the space-like hypersurface $\Sigma_t$, and reshuffling some terms, gives:
\begin{eqnarray}
\int_{\Sigma_t}\bigg[\frac{1}{2M_{p_{eff}}^2}\big(S_{i}^{\ i}-\tilde{S}_{i}^{\ i}\big)-\nu_{|i}\nu^{|i}+\frac{3}{4} \big(K_{ij}K^{ij}-K^2\big)\bigg]\sqrt{h}d^3x  =  & & \\  \nonumber
 \int_{\Sigma_t}\bigg[(Kn^{\alpha})_{;\alpha}+\nu^{|i}_{\ \ |i}-\frac{1}{4}{}^3R\bigg]\sqrt{h}d^3x. & & 
\label{eq:Step2}
\end{eqnarray}
The first term inside the integral on the RHS of~\eqref{eq:Step2} is
\begin{equation}
(Kn^{\alpha})_{;\alpha}=N^{-1}(KN^i)_{|i}=N^{-1}KN^i\nu_{|i}+(KN^i/N)_{|i}
\end{equation}
where the first equality follows from equation (2.5) of~\cite{PhysRevD.48.2635}. Therefore,
\begin{eqnarray}
& & \int_{\Sigma_t}(Kn^{\alpha})_{;\alpha}\sqrt{h}d^3x= \\ \nonumber
& & \int_{\Sigma_t}\frac{K}{N}N^i\nu_{|i}\sqrt{h}d^3x+\lim\limits_{S\rightarrow\infty}\oint_S\frac{K}{N}N^idS_i = \int_{\Sigma_t}\frac{K}{N}N^i\nu_{|i}\sqrt{h}d^3x 
\label{eq:FirstStep}
\end{eqnarray}
where the integral over the 2-surface $S$, which is diffeomorphic to a 2-sphere, goes to 0 as the radius tends to $\infty$(hence the meaning of the limit). Furthermore, the second integral of~\eqref{eq:Step2} is also a surface one, and in the limit considered, it's the total mass-energy in $\Sigma_t$~\cite{Komar:1958wp}(see appendix of~\cite{Gourgoulhon_1994} for proof):
\begin{equation}
\int_{\Sigma_t}\nu^{|i}_{\ \ |i}\sqrt{h}d^3x=\lim\limits_{S\rightarrow\infty}\oint_S\nu^{|i}dS_i=4\pi M_{\Sigma_t}.
\label{eq:KomarMass}
\end{equation}
\\The final term on the RHS of~\eqref{eq:Step2},
\begin{equation}
\int_{\Sigma_t}{}^3R\sqrt{h}d^3x
\label{eq:IntRicci}
\end{equation}
 needs to be considered carefully. The 3-Ricci scalar can be written as~\citep{lifshitz1980classical}:
\begin{equation}
{}^3R=-\frac{1}{\sqrt{h}}\frac{\partial}{\partial x^i}\bigg[\frac{1}{\sqrt{h}}\frac{\partial}{\partial x^j}(hh^{ij})\bigg]+h^{ij}\bigg[\Gamma^l_{\ \ im}\Gamma^m_{\ \ jl}-\Gamma^l_{\ \ lm}\Gamma^m_{\ \ ij}\bigg]
\label{eq:Ricci}
\end{equation}
where $\Gamma^i_{\ jk}$ are the Christoffel symbols associated with $h$. The problem is that this is not a covariant form, and its convergence into a finite value depends on the coordinate system used, as one can check by comparing~\eqref{eq:IntRicci} in spherical coordinates to its form in Cartesian ones. But all the other terms of~\eqref{eq:Step2} are indeed finite. This means that~\eqref{eq:IntRicci} must also be finite. The solution to this dilemma is to express the Ricci scalar in a form valid in any coordinate system and corresponding to the sum of a convergent surface integral and a volume integral. The latter should be written in terms quadratic in the derivative of the metric, containing only its curvature part, and not the coordinate part like the $\Gamma$s do. The key point in doing so is by introducing a flat background metric $\tilde{h}$, onto $\Sigma_t$ along with $h$. The asymptotic flatness hypothesis insures that both metrics match at infinity, and then we can write ${}^3R$ in a way covariant with respect to $\tilde{h}$. In particular, the Christoffel terms of~\eqref{eq:IntRicci} will be replaced by a quadratic term covariant with respect to $\tilde{h}$, tending to 0 in the flat-space case. This procedure is known as the bimetric formalism~\citep{cornish1964energy,nahmad1987pseudotensors,katz1990localisation}. The final form of~\eqref{eq:IntRicci} is\footnote{Note that $M_{\Sigma_t}$ in~\eqref{eq:IntRicciFinal} should be the total ADM mass energy, but because the two masses are equal in the stationary and asymptotically flat case, we skipped introducing it explicitly in the text}:
\begin{equation}
\int_{\Sigma_t}{}^3R\sqrt{h}d^3x=16\pi M_{\Sigma_t}+\int_{\Sigma_t}h^{ij}\big[\Delta^l_{\ \ im}\Delta^m_{\ \ jl}-\Delta^l_{\ \ lm}\Delta^m_{\ \ ij}\big]\sqrt{h}d^3x
\label{eq:IntRicciFinal}
\end{equation}
where
\begin{equation}
\Delta^i_{\ \ jk}\equiv\frac{1}{2}h^{il}\big[h_{lk||j}+h_{ji||k}-h_{jk||l}\big]
\end{equation}
is a covariant tensor on $\Sigma_t$, and $_{||j}$ denotes covariant derivative with respect to $x^j$ corresponding to the 3-metric $\tilde{h}.$

Ultimately, the final form of the Virial theorem in a  MGT satisfying the JBT:
\begin{eqnarray}
& & \int_{\Sigma_t}\bigg[\frac{1}{2M_{p_{eff}}^2}\big(S_i^{\ i}-\tilde{S}_i^{\ i}\big)-\nu_{|i}\nu^{|i}+  \\ \nonumber
& & \frac{1}{4}h^{ij}\big(\Delta^l_{\ \ im}\Delta^m_{\ \ jl}-\Delta^l_{\ \ lm}\Delta^m_{\ \ ij}\big)\bigg]\sqrt{h}d^3x   \\ \nonumber
& & +\int_{\Sigma_t}\bigg[\frac{3}{4}\big(K_{ij}K^{ij}-K^2\big)-\frac{K}{N}N^i\nu_{|i}\bigg]\sqrt{h}d^3x=0. 
\label{eq:FinalResult}
\end{eqnarray}
To see how this result corresponds to the known Newtonian form of the Virial theorem, consider dust particles with a stress energy tensor of the form
\begin{equation}
T^{\alpha\beta}=\rho u^{\alpha}u^{\beta}
\end{equation}
where $\rho$ is the energy density of the system and $u^{\alpha}$ is its 4-velocity vector. This means that
\begin{equation}
S_i^{\ i}= \gamma^2\rho u_iu^i
\end{equation}
where $\gamma=-n_{\alpha}u^{\alpha}$ is the Lorentz factor between the observer and the dust particles, and $u^i$ is the velocity vector measured by the observer. In the Newtonian limit, one can choose a coordinate system in which the metric becomes:
\begin{equation}
ds^2=-(1+2\nu)dt^2+(1-2\nu)\tilde{h}_{ij}dx^idx^j.
\label{eq:MetricNewt}
\end{equation}
Therefore, from~\eqref{eq:LapseVectors} and the definition of $K_{\alpha\beta}$~\citep{PhysRevD.48.2635}, one can show that $K_{ij}=K=0$. Moreover, the $\Delta\Delta$ term of the integrand becomes $1/2\nu_{||i}\nu^{||i}$ , so the net result is:
\begin{equation}
2T+\Omega-\tilde{\Omega}=0
\label{eq:VirialNewt}
\end{equation} 
where the total kinetic energy is
\begin{equation}
T\equiv\frac{1}{2}\int_{\Sigma_t}\rho u^2dV
\end{equation}
with $u^2=u^iu_i$ and $dV=\sqrt{\tilde{h}}d^3x$. Furthermore, the gravitational potential energy due to the dust particles is
\begin{equation}
\Omega\equiv-\int_{\Sigma_t}\frac{1}{M_{p_{eff}}^2}(\nabla\nu)^2dV
\end{equation}
where we put $M_{p_{eff}}^2$ inside the integral because, depending on the MGT under consideration, this term can depend on space. Finally, the additional contribution to the theorem due to the MGT is
\begin{equation}
\tilde{\Omega}\equiv\int_{\Sigma_t}\tilde{h}^{ij}\tilde{T}_{ij}dV.
\label{eq:ExtraTerm}
\end{equation}
It is clear that the latter is always present, even if one considers the systems analyzed in~\citep{Danieli_2019,van_Dokkum_2019,danieli2019tip,Guo2019wgb,Pina2019rer}. By considering these galaxies in~\eqref{eq:VirialNewt}, one can then put constrains on the parameters of the MGT considered to make~\eqref{eq:ExtraTerm} vanishingly small.

\subsection{Theories violating the JBT}
\label{Violation}

Now we consider theories, such as the Dvali-Gabadaze-Porati model~\citep{Dvali:2000hr}, which do not satisfy the JBT. Even in these theories, the equations of motion can be written in the from~\eqref{eq:Einsteinf(R)}, and hence we start our analysis from equation~\eqref{eq:Step2}.

First of all, the surface term in~\eqref{eq:FirstStep} will not go to zero, because we are no longer assuming asymptotic flatness. Furthermore, concerning the bimetric formalism trick used previously to write a covariant expression for~\eqref{eq:IntRicci}, we can still use the same analysis. However, now the metric introduced $\tilde{h}$ onto the $\Sigma_t$s in no longer flat everywhere, rather it should match the form of $h$ at infinity (which is not flat for the type of theories considered her), but remain flat where the dynamics are taking place. This is a mathematical trick to guarantee that~\eqref{eq:IntRicci} is written in a covariant way, and should not affect the physical result, specially when we take the Newtonian limit, as we will see shortly (see~\cite{cornish1964energy,nahmad1987pseudotensors,katz1990localisation} for more details). Finally, since the mass term~\eqref{eq:KomarMass}, which is known as the Komar mass~\cite{Komar:1958wp}, does not cancel the ADM one appearing in~\eqref{eq:IntRicciFinal}, the final result of the Generalized Virial theorem~\eqref{eq:FinalResult} will have three additional terms on its RHS:
\begin{equation}
\lim\limits_{S\rightarrow\infty}\oint_S\frac{K}{N}N^idS_i+4\pi\big(M_{K}-M_{ADM}\big)
\label{eq:AddTerms}
\end{equation}
where $M_K$ and $M_{ADM}$ are the Komar and ADM masses, respectively.

From here, since we are interested in studying the dynamics of galaxies and galactic clusters, we need to take the Newtonian limit, as given by the metric~\eqref{eq:MetricNewt}. In this limit, it was shown in~\cite{Abramowicz:1976nk} that the two mass terms in~\eqref{eq:AddTerms} do indeed match, and therefore cancel. Furthermore, the surface term cancels by definition from the form of the metric in this limit. Therefore, even for the case of theories that violate the JBT, in the Newtonian limit, the Virial theorem takes the form~\eqref{eq:VirialNewt}, but away from that limit it has~\eqref{eq:AddTerms} as additional terms. It should be stressed that if a MGT does not produce the Virial theorem at galactic scales, then such a theory fails in producing one of the observational proofs of the existence of DM, and therefore cannot be considered as a candidate for the latter in the first place\footnote{But such a theory might still be a viable candidate for DE.}. Moreover, the unlikeliness of MGTs violating the JBT to be DM candidates has been studied previously (see~\cite{PhysRevD.77.064016}).
\section{Observational constraints}

\begin{figure}
\centering
    \includegraphics[angle=0,clip=,width=\columnwidth]{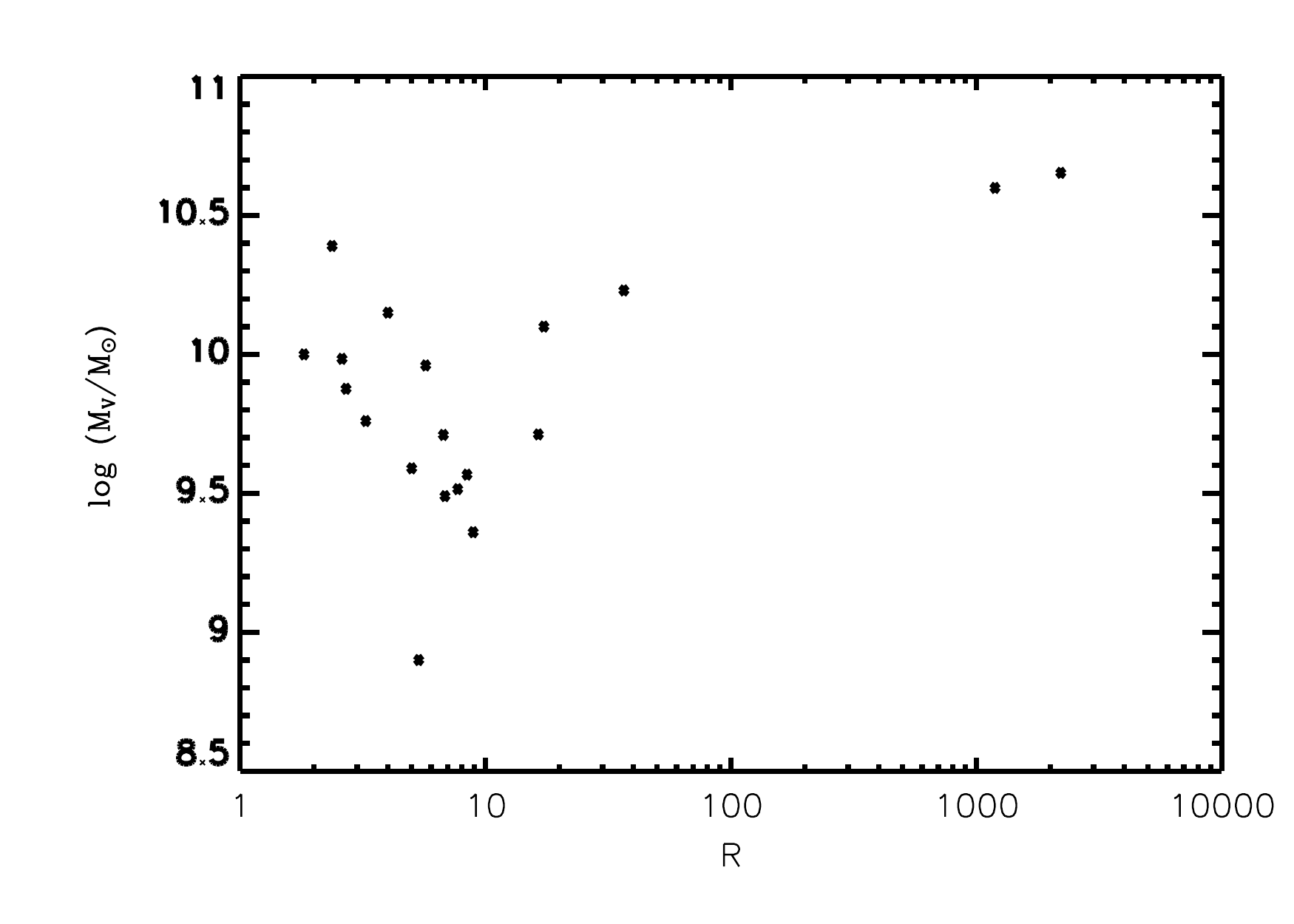}
\caption{Virial mass as a function of $R$. There is no obvious correlation or trend with the masses, which indicates modified gravity theories may need extra fine tuning as $R$ will need to be adjusted on an object by object basis.}
 \label{fig:fig1}
\end{figure}

In this section, we link the quantities obtained in the generalized Virial theorem~\eqref{eq:VirialNewt} to those that could be observed, such as in~\cite{Danieli_2019,van_Dokkum_2019,danieli2019tip,Guo2019wgb,Pina2019rer}. To this end, we can define masses and densities associated to the quantities $\Omega$ and $\tilde{\Omega}$, as written in~\eqref{Meq.} and~\eqref{Omegaeq.}. Note that the effect of MGT on $M_{p_{\rm eff}}$ has been absorbed into the masses. One can also define radii associated to the these two quantities: 
\begin{equation}
R_V=-G\frac{M^2}{\Omega},\quad R_M=\frac{G\tilde{M}^2}{\tilde{\Omega}}
\label{RadiiDef}
\end{equation}
where $R_V$ is the Virial radius and $R_M$ is the radius in which the MGT takes effect. According to~\cite{10.1093/mnras/148.3.249}, the Virial mass is defined as 
\begin{equation}
2T=\frac{GM_V^2}{R_V}
\end{equation}
which, when inserted in~\eqref{eq:VirialNewt}, with the use of~\eqref{RadiiDef}, gives the following relation between the Virial, baryonic and MGT masses:
\begin{equation}
\frac{M_V^2}{M^2}=1+\frac{\tilde{M}^2R_V}{M^2R_M}
\label{MassToRadiusConst.}
\end{equation}
and therefore, one way of constraining the ratio $\tilde{M}^2/R_M$, i.e a constrain on $\tilde{\Omega}$, is by measuring $M,M_V$ and $R_V$ and using~\eqref{MassToRadiusConst.}. In table~\ref{table:1} we present the constraint on the ratio $\tilde{\Omega}/\Omega$ given by the measurements presented in~\cite{Guo2019wgb}. This ratio is calculated using equations~\eqref{RadiiDef} inserted in~\eqref{MassToRadiusConst.}. As we can see in table~\ref{table:1} and Fig.~\ref{fig:fig1}, $\tilde{\Omega}$ is an appreciable multiple of $\Omega$, when in fact, for these DM devoid galaxies, it should be negligible. This shows that, unless another mechanism is introduced specially for these galaxies to remove $\tilde{\Omega}$, it is very difficult for MGTs to account for these galaxies and for other DM rich ones that are similar in properties, such as AGC 8915, for instance~\cite{Guo2019wgb}. 
\begin{table}
\caption{
Data from~\citet{Guo2019wgb}.  
The second and third columns are the logarithm of the baryonic and Virial masses, respectively. The ratio $|\tilde{\Omega}/\Omega|$ shows no obvious correlation or trend with the masses, which indicates modified gravity theories may need extra fine tuning as $R$ will need to be adjusted on an object by object basis.}
	\centering
	\begin{tabular}{cccc} 
		\hline
		Galaxy Name & $\log(M_b/M_{\odot}$) & $\log(M_V/M_{\odot}$) & $R=|\tilde{\Omega}/\Omega|$ \\ 
		\hline
		AGC 6438 & 9.444 & 10.231 & 36.497 \\ 
		AGC 6980 & 9.592 & 9.876 & 2.698 \\
		AGC 7817 & 9.061 & 10.599 & 1190.242 \\
		AGC 7920 & 8.981 & 10.653 & 2207.005 \\
		AGC 7983 & 9.046 & 9.515 & 7.700 \\ 
		AGC 9500 & 9.092 & 9.712 & 16.378\\
		AGC 191707 & 9.080 & 9.567 & 8.419 \\
		AGC 205215 & 9.706 & 9.984 & 2.597 \\
		AGC 213086 & 9.8 & 10.149 & 4.000 \\
		AGC 220901 & 8.864 & 9.363 & 8.954 \\
		AGC 241266 & 9.547 & 9.96 & 5.699 \\
		AGC 242440 & 9.467 & 10.098 & 17.281 \\
		AGC 258421 & 10.124 & 10.387 & 2.373 \\
		AGC 321435 & 9.204 & 9.593 & 4.998 \\
		AGC 331776 & 8.503 & 8.904 & 5.339 \\
		AGC 733302 & 9.042 & 9.489 & 6.834 \\
		AGC 749244 & 9.778 & 10.003 & 1.818 \\
		AGC 749445 & 9.264 & 9.708 & 6.727 \\
		AGC 749457 & 9.445 & 9.759 & 3.246 \\[1ex] 
		\hline
	\end{tabular}
		\label{table:1}
\end{table}

In addition to that, another observable parameter that can be used is the velocity dispersion, $\sigma$, related to the Virial mass and radius by~\cite{Munari_2013,10.1093/mnras/stx562}:
\begin{equation}
M_V=\frac{3\sigma^2R_V}{G}.
\label{Dispersion}
\end{equation}
 This can be used with~\eqref{MassToRadiusConst.} to put another constraint on a given MGT. Indeed, we can write~\eqref{MassToRadiusConst.} as:
\begin{equation}
\frac{\sigma_g^4}{\sigma_{\text{int}}^4}=1+\frac{\tilde{M}^2R_V}{M^2R_M}
\end{equation}
where $\sigma_{\text{int}}$ and $\sigma_g$ are the intrinsic and globular cluster's velocity dispersions, respectively. For instance, if we use the results of~\cite{van_Dokkum_2019}, where the velocity dispersion of DM devoid galaxy NGC 1052-DF2 has been measured, we find 
\begin{equation}
\frac{\tilde{M}^2}{R_M}\approx 3\frac{M^2}{R_V}\Rightarrow \tilde{\Omega}\approx 3\Omega.
\end{equation}
In other words, if DM is described by a MGT, the latter should produce a gravitational potential approximately 3 times that of baryonic matter for NGC 1052-DF2, where in reality it should not be present. 

A third method to check the consequences of MGTs on DM devoid galaxies is by writing the mass in a radius $r$ of a system as:
\begin{equation}
M(r)=4\pi\int_{0}^{r}\big(\rho_B+\tilde{\rho}\big)r'^2dr'=M_B+\tilde{M}
\end{equation}
where $\rho_B$ and $M_B$ are the density and mass of baryonic matter, respectively, while, $\tilde{\rho}$ and $\tilde{M}$ are those of MGT. By measuring $M(r)$ and $M_B$ one can therefore determine the amount of DM available as a MGT. Note that this is independent of whether a system is virialzed or not. For example, in~\cite{Danieli_2019} the dynamical mass and that of the stars within the half-light radius for the ultra diffuse galaxy NGC1052-DF2 have been measured to be very similar. This puts dire constraint on MGTs, which highlights the importance of such DM devoid galaxies in constraining these theories.
\section{Conclusions}
\label{sec:conclusions}

We have computed the Virial theorem for MGTs that satisfy the JBT, as well as those that don't. Motivated by the recent discovery of a class of dwarf galaxies with no significant DM, we wanted to quantify what constraints these objects put on MGTs. In the same vein that the number of satellite dark matter halos imposes sever constraints on the nature of particle dark matter, we have found an equivalent observable for the case when dark matter is a modification of gravity.  Inspection of \eqref{eq:FinalResult} and~\eqref{eq:VirialNewt} shows that the Virial theorem for MGT contains an extra term $\tilde{S}_i^i$ or $\tilde{\Omega}$. The existence of this term can be constrained by the DM devoid galaxies considered here. For instance, it was shown in~\cite{Capozziello_2013} that this extra mass term is proportional to the baryonic mass present in the system. If one then applies this model to the galaxies at hand, as is presented in table~\ref{table:1}, it would be difficult to see how the model matches the observations without fine tuning. On the other hand, trying to accommodate this term for MGT models will provide interesting insights into the nature of these models. If DM can indeed be part of the theory of gravity, one can think of two possibilities that $\tilde{S}_i^{\ i}$ can have in order to achieve that. The first is that $\tilde{S}_i^{\ i}$ should include specific coupling terms dependent on the environment and baryonic content of the considered galaxies in~\cite{Danieli_2019,van_Dokkum_2019,danieli2019tip,Guo2019wgb,Pina2019rer} such that they cancel the terms that generate DM effects in other galaxies. That is, the matter content and configuration of these galaxies should couple to gravity in a special way in order to make sure there's no DM effect. But this puts the universality of gravitational interactions into question. So another way is to look at a map of the sky for DM distribution, and have $\tilde{S}_i^{\ i}$ be the function that goes to 0 at the special positions where these galaxies are found, while it is not 0 in other locations. However, the difficulty arises form the fact that most dwarf galaxies do have dark matter, in fact are dark matter dominated, which makes the above suggested solution highly fine-tuned. On the other hand, if DM was some type of particles, then accommodating its absence in these galaxies would be less fine tuned, by using, for instance, hydrodynamical events associated with galaxy formation. It will be interesting to see if non-fine tuned MGT can be constructed to fulfill the existence of dark-matter-free galaxies~\citep{Danieli_2019,van_Dokkum_2019,danieli2019tip}.


\section*{Acknowledgments}

We would like to thank Nicola Bellomo for interesting and fruitful discussions. We also thank the anonymous referee for a very positive and useful report. Funding for this work was partially provided by the Spanish science ministry under project PGC2018-098866-B-I00. 

\section*{Data Availability}
The data underlying this article are available in the article and references therein.


\appendix
\section{Alternative Derivation of the Virial Theorem}\label{Appendix}

We present here another method for deriving the Virial Theorem~\eqref{eq:VirialNewt}, using the Lagrangian formalism and the relativistic Boltzmann equation. The method presented here should be applicable to any metric gravity theory, since it follows from energy-momentum conservation. Therefore it applies to theories that violate the JBT as well. First, consider the equations of motion~\eqref{eq:Einsteinf(R)} and the metric in spherical coordinates:
\begin{equation}
ds^2=-e^{2\nu(r)}dt^2+e^{-2\nu(r)}dr^2+r^2(d\theta^2+\sin{\theta}^2d\phi^2)
\end{equation}
(at the moment no approximations are being made. When we apply the Newtonian approximation, this metric reduces to~\eqref{eq:MetricNewt}). The $0-0$, $r-r$, $\theta-\theta$ and $\phi-\phi$ components of the field equations are, respectively:
\begin{equation}
-e^{2\nu}\bigg(\frac{1}{r^2}+\frac{2\nu'}{r}\bigg)+\frac{1}{r^2}=\frac{1}{M_{p_{eff}}^2}\big(T_{00}+\tilde{T}_{00}\big)
\end{equation}
\begin{equation}
e^{2\nu}\bigg(\frac{2\nu'}{r}+\frac{1}{r^2}\bigg)-\frac{1}{r^2}=\frac{1}{M_{p_{eff}}^2}\big(T_{rr}+\tilde{T}_{rr}\big)
\end{equation}
\begin{equation}
\frac{1}{2}e^{2\nu}\bigg(2\nu''+4\nu'^2+4\frac{\nu'}{r}\bigg)=\frac{1}{M_{p_{eff}}^2}\big(T_{\theta\theta}+\tilde{T}_{\theta\theta}\big)
\end{equation}
\begin{equation}
\frac{1}{2}e^{2\nu}\bigg(2\nu''+4\nu'^2+4\frac{\nu'}{r}\bigg)=\frac{1}{M_{p_{eff}}^2}\big(T_{\phi\phi}+\tilde{T}_{\phi\phi}\big).
\end{equation}
Summing these equations together, we get:
\begin{equation}
e^{2\nu}\bigg(2\nu''+\frac{4\nu'}{r}+4\nu'^2\bigg)=\frac{1}{M_{p_{eff}}^2}\big(T_{tot}+\tilde{T}_{tot}\big)
\label{eq:Sum}
\end{equation}
where $T_{tot}$ and $\tilde{T}_{tot}$ are the sum of the components of $T$ and $\tilde{T}$, respectively. Assuming that the deviation from GR is small, one can write $M_{p_{eff}}^2=M_p^2(1+\epsilon\Psi)$ where $\epsilon$ is a small quantity and $\Psi$ describes the deviation from GR due to the presence of $\tilde{T}_{\mu\nu}$. Equation~\eqref{eq:Sum} becomes:
\begin{equation}
e^{2\nu}\big(2\nu''+\frac{4\nu'}{r}+4\nu'^2\big)=\frac{1}{M_{p}^2}\big(T_{tot}+2\tilde{\rho}\big)
\label{eq:SumFinal}
\end{equation}
and $2\tilde{\rho}=\tilde{T}_{tot}(1-\epsilon\Psi)$, written in this form for later convenience.

Next step, consider a system of collisionless point particles following a distribution function $f_B$. The stress energy tensor of such a system can be defined as:
\begin{equation}
T_{\mu\nu}=\int f_B mu_{\mu}u_{\nu}du
\end{equation}
where $m$ is the mass of a particle (galaxy, star...), $u_{\mu}$ its 4-velocity, and $du=du_rdu_{\theta}du_{\phi}/u_t$ the invariant volume element in velocity space. From this definition, one can write
\begin{equation}
\frac{1}{M_p^2}T_{tot}=\frac{2}{M_p^2}\rho\langle u^2\rangle
\label{eq:Ttot}
\end{equation}
where $\rho$ is the mass density of the system, and $\langle u^2\rangle =\langle u^2_t\rangle+\langle u^2_r\rangle+\langle u^2_{\theta}\rangle+\langle u^2_{\phi}\rangle$, with $\langle.\rangle$ being the average in velocity space. The distribution function $f_B$ follows the relativistic Boltzmann equation~\cite{Bildhauer_1989,Maartens:1985dn}:
\begin{equation}
\bigg(p^{\alpha}\frac{\partial}{\partial x^{\alpha}}-p^{\alpha}p^{\beta}\Gamma^{i}_{\ \alpha\beta}\frac{\partial}{\partial p^i}\bigg)f_B=0
\label{eq:RelativisticBoltzman}
\end{equation}
where $p^{\alpha}$ is the particle's 4-momentum (see~\cite{Bohmer:2007fh} for further mathematical details). At this stage, it is more convenient to introduce a set of local tetrads $e^a_{\mu}(x), a=0,1,2,3$, which can be chosen to be, for the current case of spherical symmetry:
\begin{align}
&e^0_{\mu}=e^{\nu}\delta^0_{\mu},\quad e^1_{\mu}=e^{-\nu}\delta^1_{\mu}
\\
&e^2_{\mu}=r\delta^2_{\mu},\quad e^3_{\mu}=r\sin{\theta}\delta^3_{\mu}
\end{align}
where $\delta^a_{\mu}$ is the Kronecker delta. Assuming that $f_B=f_B(r,u^a)$, where $u^a=u^{\mu}e^a_{\mu}$ are the velocity components in the tetrad frame, equation~\eqref{eq:RelativisticBoltzman} becomes~\cite{10.1093/mnras/148.3.249}:
\begin{eqnarray}
& & u_1\frac{\partial f_B}{\partial r}-\bigg(u_0^2\frac{\partial\nu}{\partial r}-\frac{u_2^2+u_3^2}{r}\bigg)\frac{\partial f_B}{\partial u_1} \\ \nonumber
& & -\frac{1}{r}u_1\bigg(u_2\frac{\partial f_B}{\partial u_2}+u_3\frac{\partial f_B}{\partial u_3}\bigg) \\ \nonumber 
& & -\frac{1}{r}e^{-\nu}u_3\cot{\theta}\bigg(u_2\frac{\partial f_B}{\partial u_3}-u_3\frac{\partial f_B}{\partial u_2}\bigg)=0.
\end{eqnarray}
Multiplying the above equation by $mu_rdu$ and integrating over the velocity space (assuming that $f_B\rightarrow0$ as $u\rightarrow\pm\infty$), then multiplying by $4\pi r^2dr$ and integrating over the system, we get finally:
\begin{align}
\int_{0}^{R}4\pi\rho&\big[\langle u_1^2\rangle+\langle u_2^2\rangle+\langle u_3^2\rangle\big]r^2dr\nonumber\\&-\frac{1}{2}\int_{0}^R4\pi\rho\big[\langle u_0^2\rangle+\langle u_1^2\rangle\big]r^3\frac{\partial\nu}{\partial r}dr=0.
\label{eq:BoltzmanFinal}
\end{align}
To simplify the problem, one can make two further approximations. First, assume $\nu$ to be small and slowly varying, hence $e^{2\nu}\approx1+2\nu $ and all quadratic terms in $\nu$ or $\nu'$ drop. Second, assume the velocities to be much smaller than the speed of light, therefore $ \langle u_1^2\rangle\approx \langle u_2^2\rangle \approx\langle u_3^2\rangle\ll\langle u_0^2\rangle\approx1$. Thus, eqs.~\eqref{eq:SumFinal}(after using~\eqref{eq:Ttot}) and~\eqref{eq:BoltzmanFinal} become:
\begin{equation}
\frac{1}{r^2}\frac{\partial}{\partial r}\bigg(r^2\frac{\partial\nu}{\partial r}\bigg)=\frac{1}{M_p^2}(\rho+\tilde{\rho})
\label{eq:SumFinal2}
\end{equation}
and
\begin{equation}
2T-\frac{1}{2}\int_0^R4\pi\rho\frac{\partial \nu}{\partial r}r^3dr=0
\label{eq:Kinetic}
\end{equation}
respectively, where
\begin{equation}
T=\int_0^R2\pi\rho\big[\langle u_1^2\rangle+\langle u_2^2\rangle+\langle u_3^2\rangle\big]r^2dr
\end{equation}
is the total kinetic energy of the system. Multiplying~\eqref{eq:SumFinal2} by $r^2$ and integrating from $0$ to $r$, we get, when using the explicit form of $M_p^2$ given in section~\ref{sec:intro}:
\begin{equation}
GM(r)=\frac{1}{2}r^2\frac{\partial\nu}{\partial r}-G\tilde{M}(r)
\label{eq:BeforeLastStep}
\end{equation}
where
\begin{equation}
M(r)\big(\tilde{M}(r)\big)=4\pi\int_0^R\rho\big(\tilde{\rho}\big)r'^2dr'.
\label{Meq.}
\end{equation}
Finally, multiplying~\eqref{eq:BeforeLastStep} by $dM(r)$ and integrating from $0$ to $R$, after the use of~\eqref{eq:Kinetic}, we get the generalized Virial theorem:
\begin{equation}
2T+\Omega+\tilde{\Omega}=0
\end{equation}
where
\begin{equation}
\Omega\big(\tilde{\Omega}\big)=-\int_0^R\frac{GM(r)\big(\tilde{M}(r)\big)}{r}dM(r).
\label{Omegaeq.}
\end{equation}

\end{document}